\documentstyle[aps,psfig]{revtex}
\begin{document}
\draft

\author{J. P. Dewitz, W. H\"ubner, and K. H. Bennemann}
\address{Institute for Theoretical Physics, Freie Universit\"at
  Berlin, Arnimallee 14, 14195 Berlin, Germany.}
\date{\today}
\title{Theory for the nonlinear optical response of a nonspherical
  metal cluster}
\maketitle
\begin{abstract}
Using classical electrodynamics we determine the higher harmonic
radiation by a nonspherical metal cluster in form of a matrix
formalism. Extending the theory for the source of the higher harmonic
radiation for spherical metal clusters as introduced by \"Ostling et
al. [Z. Phys. D
{\bf 28}, 169 (1993)] we calculate the sources for nonspherical
particles. Employing these sources we develop the nonlinear Mie theory and
determine the radiated higher harmonic fields generated by the
cluster. Our theory is valid for arbitrary shape and arbitrary complex
refractive index for cluster sizes much smaller and comparable to the
wavelength of the incident light.
\end{abstract}
\section{Introduction}
The scattering behavior of
clusters can be divided approximately into two regimes. First the scattering
by clusters with sizes much smaller than the wavelength of light, the
classical Rayleigh limit, which is dominated by microscopic effects
and second the Mie range when the particle size becomes equal to or larger than
the wavelength of light. The latter is characterized by resonances and
interferences of surface electromagnetic modes described by classical
physics. Since we are interested in this regime, we neglect
microscopic phenomena. For {\em spherical} particles, the macroscopic effects
known from linear Mie scattering~\cite{BorWol,BohHuf} are even more pronounced
in second and third harmonic generation, SHG and THG,
respectively. Furthermore, they depend much more sensitively on the
size and refractive index in the higher harmonic case~\cite{Dew}. For
{\em nonspherical} particles, another important
parameter governs the value and the angular dependence of the radiated
intensities, viz. the curvature. Apart from the problem of determining
this parameter as
well as the size and refractive index from the angular dependence of
the scattered intensities, the influence of clusters or well-defined
nanostructures at surfaces on the SHG signal in reflection, especially
the size dependence and the absolute value of the nonlinear
response~\cite{Goe} are of particular interest.
An important advantage for the application of higher harmonic
techniques is that they are, in contrast to linear optics, not
influenced by background noise.
Knowledge of the intensities radiated by
nanoparticles deposited on surfaces or by surface roughnesses of
nanometer size, should be one ingredient to determine the lateral
spatial resolution of SHG theoretically, the lower limit of which is as low is
as 1~nm~\cite{Akt} due to local field enhancements. Even in this range,
the classical Rayleigh limit, an essential contribution results from
the intensities radiated by {\em nonspherical} particles.

In the literature one can find several approximate theories for the
linear radiation from nonspherical clusters. The
Rayleigh-Gans theory is valid for arbitrary shapes but only for
materials where the absolute value of the complex refractive
index is close to one ($|\epsilon|\simeq1$). Latimer~\cite{Lat} calculates the
intensities scattered from ellipsoidal particles by relating every single point
on the surface to an equivalent sphere and determines their linear
response within Mie theory~\cite{Mie}. Other theories determine the
intensities from aspherical particles of arbitrary size within a
perturbation approach in terms of the linear deviations from the
spherical shapes~\cite{Elw}. In
the special case of spheroidal particles Asano and
Yamamoto~\cite{AsaYam} use spheroidal coordinates which decouple the
vector wave equations to obtain the exact results.

Here we extend our theory for the nonlinear response of a
spherical particle~\cite{Dew,Sta} to a particle of arbitrary
shape. Therein the sources of the higher
harmonic radiation are treated in terms of the surface charge $\sigma$,
which is, in the case of metals, equal to the normal projection of the
polarization on the the surface ($\sigma={\bf n}\cdot{\bf P}$) and
approximates the induced surface charge in the $n$-th harmonic case by
the $n$-th power of the linear induced surface charge
$\sigma^{(n)}\equiv(\sigma^{(1)})^n$. This approach is motivated by
the classical anharmonic oscillator model (see~\cite{Dew}) which is
valid in the Mie regime and gives a good approximation for particles
with can be considered as locally smooth. The $n$-th power of the
surface charge,
which is induced by the incident field acts as the source for the
higher harmonic fields. It yields a discontinuity of the electrical
displacement at the surface which can be expressed in terms of the
boundary conditions for the radial component of the electrical
fields. Using also the continuity of the tangential component of the
electric field, the radiated fields can be fully determined. This is a
simple task for spherical particles if spherical coordinates are
used. In the case of nonspherical particles, however,
matrix equations come into play where the size of the matrices depends
on the ratio of the particle dimension to the wavelength of the light
and also the geometry of the shape.

To obtain the results for the linear scattering, which are necessary
to determine the induced surface charge $\sigma^{(1)}$ a formalism by
Barber and Yeh~\cite{BarYeh} is used. This is based on an
integral equation method by Waterman~\cite{Wat1,Wat2} and
Schelkunoff's equivalence principle~\cite{Sch} yielding matrix
equations. This is a good choice because Barber and Yeh use nearly the same
representation of the fields with spherical coordinates and their
theory is valid for arbitrary cluster-shapes and arbitrary complex
indices of refraction.
\section{theory}
We determine the $n$-th harmonic electric field ${\bf E}^{(n)}_{\rm
  out}$ radiated
from arbitrarily shaped particles as a function of the $n$-th power of
the linear
surface charge $\sigma^{(1)}$ induced by the incident field which acts as
the source of the $n$-th harmonic. The linear surface charge is given
by $\sigma^{(1)}={\bf
  n}\cdot{\bf P}^{(1)}$ where
{\bf n} is the vector normal at the surface of the cluster since in
the case of metals only the normal component of the polarization is
nonzero. The source $\sigma^{(n)}=(\sigma^{(1)})^n$ induces a
discontinuity of the
normal component of the dielectrical displacement at the boundary of the
particle. As a consequence we use the conventional boundary conditions for the
components of the electric fields to derive ${\bf E}^{(n)}_{\rm
  out}$ which then determines the scattering profile and all the
characteristics of the nonlinear Mie scattering by nonspherical
particles. Obviously
\[{\bf n}\cdot\left({\bf D}_{\rm out}^{(n)}-{\bf D}_{\rm
    in}^{(n)}\right)=4\pi\sigma^{(n)}\]
and
\[{\bf n}\times\left({\bf E}_{\rm out}^{(n)}-{\bf E}_{\rm
    in}^{(n)}\right)=0\;,\]
together with ${\bf D}=\epsilon{\bf E}$, yield ${\bf E}^{(n)}_{\rm
  out}$ when $\sigma^{(n)}$ is given.

For the calculation all quantities, the fields and the source, are
expanded in terms of spherical harmonics. This results in a matrix
equation for the coefficient vectors of the fields and the source
which completely determines ${\bf
  E}^{(n)}_{\rm out}$. We show in the following how this analysis can
be performed in particular for the nonlinear Mie scattering of
nonspherical particles, such an analysis had not been done before.
\subsection{Determination of the source $\sigma^{(n)}$}
To determine the source $\sigma^{(n)}$ we use the approximation
$\sigma^{(n)}=(\sigma^{(1)})^n$. To obtain the $n$-th power of the
linear charge we proceed as follows.
The linear surface charge is defined as
\begin{equation}
\sigma^{(1)}\left(\theta,\varphi\right)={\bf n}\cdot{\bf
  P}^{(1)}\left(\theta,\varphi\right)\;.
\label{sigmaP}
\end{equation}
and can be expressed as
\begin{equation}
\sigma^{(1)}\left(\theta,\varphi\right)=\frac{1}{4\pi}{\rm
  Re}\left[\left(1-\epsilon^{-1}\right)\left({\bf E}_{\rm sc}+
{\bf E}_{\rm inc}\right)\cdot{\bf n}\right]{\rm e}^{-{\rm i}\omega t}\;.
\label{sigmaein}
\end{equation}
Here, we denote the scattered field by the index ``sc'' and the incident field
by ``inc''.
Therein we refer to the boundary conditions of the normal component of the
dielectrical displacement in the linear case ${\bf
  n}\cdot({\bf D}_{\rm sc}+{\bf D}_{\rm inc})={\bf n}\cdot{\bf D}_{\rm
    in}$ (the index ``in'' states the field inside the cluster) and to
  the relation ${\bf
  D}=\epsilon{\bf E}={\bf E}+4\pi{\bf P}$.
Furthermore, $\sigma^{(1)}$ is expanded in terms of spherical harmonics
\begin{equation}
\sigma^{(1)}\left(\theta,\varphi\right)=\frac{1}{2}\sum_{l,m=\pm1}a_{l,m}^{
\left(1\right)}Y_{l,m}\left(\theta,\varphi\right){\rm e}^{-{\rm
    i}\omega t}+c.c.\;.
\label{sigma^1}
\end{equation}
{}From this it is easy to expand the $n$-th power of the linear
surface charge in the form
\begin{equation}
(\sigma^{(1)})^n\left(\theta,\varphi\right)=\frac{1}{2}\sum_{l,m}a_{l,m}^{
\left(n\right)}
Y_{l,m}\left(\theta,\varphi\right){\rm e}^{-{\rm i}n\omega t}+c.c.\;.
\label{sigma1^n}
\end{equation}
by raising the linear surface charge $\sigma^{(1)}$ to the $n$-th power
and neglecting terms with time-dependence different from ${\rm
  e}^{{\rm i}n\omega t}$. Thus, we get the nonlinear sources $\sigma^{(n)}$.
 As an example the coefficients of $\sigma^{(2)}$ yield (\cite{Dew,Sta})
\begin{eqnarray}
a_{l,2}^{\left(2\right)}&=&\frac{1}{2}\sum_{l_1=1}^{\infty}\sum_{l_2=1}^{
\infty}a_{l_1,1}^{\left(1\right)}a_{l_2,1}^{\left(1\right)}\int
Y_{l,2}^*Y_{l_1,1}
Y_{l_2,1}d\Omega\;,\nonumber\\
a_{l,-2}^{\left(2\right)}&=&\frac{1}{2}\sum_{l_1=1}^{\infty}\sum_{l_2=1}^{
\infty}a_{l_1,-1}^{\left(1\right)}a_{l_2,-1}^{\left(1\right)}\int
Y_{l,-2}^*Y_{l_1,-1}Y_{l_2,-1}d\Omega\;,\nonumber\\
a_{l,0}^{\left(2\right)}&=&\frac{1}{2}\sum_{l_1=1}^{\infty}\sum_{l_2=1}^{
\infty}a_{l_1,1}^{\left(1\right)}a_{l_2,-1}^{\left(1\right)}\int
Y_{l,0}^*Y_{l_1,1}Y_{l_2,-1}d\Omega\;.
\label{al2}
\end{eqnarray}
Therein the integrals over the Legendre Polynomials $Y_{l,m}$ can be
expressed in terms of the well known 3$j$-symbols.
To calculate the coefficients the fields ${\bf E}_{\rm out}$ and ${\bf
  E}_{\rm sc}$, which enter via Eq.~\ref{sigmaein}, are expanded
in terms of vector
spherical harmonics as introduced by Jackson~\cite{Jac} in the form
\begin{eqnarray}
 {\bf E}_i^{(n)}\left({\bf
     x}\right)&=&\sum_{l,m}C\left(l\right)\left[K_{\rm
     M}\left(l,m\right)f^i_l
\left(k_1r\right){\bf X}_{l,m}\left(\theta,\varphi\right)\right.\nonumber\\
&
&\left.\qquad\qquad\qquad\qquad\qquad+\frac{m}{\left|m\right|}K_{\rm
    E}\left(l,m\right)\frac{1}{\epsilon\left(n\omega\right)k}\vec{\nabla}\times
f^i_l\left(k_1r\right){\bf X}_{l,m}\left(\theta,\varphi\right)\right]\;.
\label{Eent}
\end{eqnarray}
Therein, ${\bf
  X}_{l,m}(\theta,\varphi)={\bf
  L}Y_{l,m}(\theta,\varphi)/\sqrt{l(l+1)}$ is a vector spherical harmonic
({\bf L}$=1/{\rm i}({\bf r}\times\vec{\nabla})$
is the angular momentum operator),
$C(l)={\rm i}^l\sqrt{4\pi(2l+1)},k=n\omega/c$ and
$k_1=\sqrt{\epsilon(n\omega)}k$. The multipole coefficients $K
_{\rm M}^{(n)}(l,m)$ and $K_{\rm E}^{(n)}(l,m)$ refer to the {\em magnetic}
(transverse electric TE) and {\em electric} (transverse magnetic TM)
multipoles.
The index
$i$ specifies the fields external ($i\equiv$out) or internal
($i\equiv$in) to the cluster. The
spherical Hankel functions
$f^{\rm out}_l(kr)=h_l^{(1)}(kr)$ and
Bessel functions $f^{\rm in}_l(kr)=j_l(k_1r)$
describe the normal projection of the field inside and outside the
particle. In the linear case the external field splits into the
incident and the scattered field ${\bf E}_{\rm inc}$ and ${\bf E}_{\rm
  sc}$, respectively.
The coefficients of ${\bf E}_{\rm inc}$ are known from the input field
while ${\bf E}_{\rm sc}$ can be
calculated by using the results from the theory for the linear
problem by Barber and Yeh~\cite{BarYeh}. They expressed the
coefficients of the internal field in terms of the coefficients of the
incident field and then obtained the coefficients of the external
field in terms
of the coefficients of the internal field.

In higher harmonic radiation where we will use the same expansion of
the fields, however, only the
external field ${\bf E}_{\rm out}^{(n)}$ radiated from the oscillating
surface charge and the internal field ${\bf E}_{\rm in}^{(n)}$ occurs,
since no incident field is
present.
\subsection{Determination of the radiated field ${\bf E}^{(n)}_{\rm out}$}
To calculate the radiated field ${\bf E}_{\rm out}^{(n)}$ it is more
convenient to introduce the abbreviations
\begin{equation}
  \label{abbrevK}
  K'_{\rm M}(l,m)\equiv C(l)K_{\rm M}(l,m)\quad,\quad
  K'_E(l,m)\equiv\frac{m}{|m|}\frac{1}{\epsilon(\omega)k}C(l)K_E(l,m)
\end{equation}
in the sum representation of Eq.~\ref{Eent}. This leads to
\begin{equation}
  \label{EFeldkurz}
  {\bf E}_i^{(n)}\left({\bf x}\right)=\sum_{l,m}\left[K'_{\rm
      M}(l,m)f^i(k_1r){\bf
      X}_{l,m}\left(\theta,\varphi\right)+K'_E(l,m)\vec{\nabla}\times
    f_l(k_1r){\bf X}_{l,m}(\theta,\varphi)\right]\;.
\end{equation}
In the $n$-th harmonic case the source $\sigma^{(n)}$
acts as the discontinuity of the normal part of the electrical displacement
\begin{equation}
{\bf n}\cdot\left({\bf D}_{\rm out}^{(n)}-{\bf D}_{\rm
  in}^{(n)}\right)=4\pi\sigma^{(n)}\;.
\label{nD=sigma}
\end{equation}
(according to our model for the nonlinear sources $\sigma^{(n)}$
equals $(\sigma^{(1)})^n$). Furthermore we will use the continuity
condition of the
tangential component of the electric fields
\begin{equation}
\label{nxE=0}
{\bf n}\times\left({\bf E}_{\rm out}^{(n)}-{\bf E}_{\rm
    in}^{(n)}\right)=0\;.
\end{equation}
For particles of arbitrary shape the radius vector ${\bf e}_{\rm r}$
is no longer parallel to {\bf n} and the radius $r$ rather becomes a
function of
$\theta$ and $\varphi$ at the boundary since every point of the surface
is determined by the angles. Furthermore the normal vector {\bf n} now has
the form
\[{\bf n}(\theta,\phi)={\rm n}_{\bf r}(\theta,\varphi)\cdot{\bf e}_{\rm r}+{\rm
  n}_\theta(\theta,\varphi)\cdot{\bf e}_\theta+{\rm
  n}_\varphi(\theta,\varphi)\cdot{\bf e}_\varphi\;.\]
Here, $({\bf e}_{\rm r},{\bf e}_\theta,{\bf e}_\varphi)$ are the basis
vectors corresponding to spherical coordinates. As a result
Eq.~\ref{nxE=0} splits into three equations, one for every component
\begin{eqnarray}
  \label{eqcomp}
{\bf n}\cdot\left({\bf D}^{(n)}_{\rm out}-{\bf D}^{(n)}_{\rm
    in}\right)&=&4\pi\sigma^{(n)}\nonumber\\
{\bf e}_{\rm r}\cdot\left[{\bf n}\times\left({\bf E}^{(n)}_{\rm out}-{\bf
    E}^{(n)}_{\rm in}\right)\right]&=&0\nonumber\\
{\bf e}_\theta\cdot\left[{\bf n}\times\left({\bf E}^{(n)}_{\rm out}-{\bf
    E}^{(n)}_{\rm in}\right)\right]&=&0\nonumber\\
{\bf e}_\varphi\cdot\left[{\bf n}\times\left({\bf E}^{(n)}_{\rm out}-{\bf
    E}^{(n)}_{\rm in}\right)\right]&=&0\;.
\end{eqnarray}
By using the series representations
\begin{eqnarray}
  \label{EReihen}
  {\bf E}_{\rm out}^{(n)}&\equiv&\sum_{l,m}\left[A_{\rm
    E}^{(n)}h_l^{(1)}(kr){\bf
      X}_{l,m}+A_{\rm M}^{(n)}\vec{\nabla}\times h^{(1)}_l(kr){\bf
      X}_{l,m}\right]\nonumber\\
  {\bf E}_{\rm in}^{(n)}&\equiv&\sum_{l,m}\left[B^{(n)}_{\rm E}j_l(k_1r){\bf
      X}_{l,m}+B^{(n)}_{\rm M}\vec{\nabla}\times j_l(k_1r){\bf
      X}_{l,m}\right]
\end{eqnarray}
multiplying every equation by $Y_{l',m'}$ where $(l',m')$ runs over
all $(l,m)$-combinations used in the series representations of the
fields and the source and integrating over the solid angle we get the
following matrix equations
\begin{eqnarray}
  \label{MatrixGl}
  {\sf S}^1{\bf A}^{(n)}_{\rm M}+{\sf S}^2{\bf A}^{(n)}_{\rm E}-{\sf
    S}^3{\bf B}^{(n)}_{\rm M}-{\sf S}^4{\bf B}^{(n)}_{\rm
    E}&=&2\pi{\bf a}^{(n)}\nonumber\\
{\sf M}_{\rm r}^1{\bf A}_{\rm M}^{(n)}+{\sf M}_{\rm r}^2{\bf A}_{\rm
  E}^{(n)}-{\sf M}_{\rm r}^3{\bf B}_{\rm M}^{(n)}-{\sf M}_{\rm
  r}^4{\bf B}_{\rm E}^{(n)}&=&0\nonumber\\
{\sf M}_\theta^1{\bf A}_{\rm M}^{(n)}+{\sf M}_\theta^2{\bf A}_{\rm
  E}^{(n)}-{\sf M}_\theta^3{\bf B}_{\rm M}^{(n)}-{\sf M}_\theta^4{\bf
  B}_{\rm E}^{(n)}&=&0\nonumber\\
{\sf M}_\varphi^1{\bf A}_{\rm M}^{(n)}+{\sf M}_\varphi^2{\bf A}_{\rm
  E}^{(n)}-{\sf M}_\varphi^3{\bf B}_{\rm M}^{(n)}-{\sf M}_\varphi^4{\bf
  B}_{\rm E}^{(n)}&=&0\;.
\end{eqnarray}
The coefficients are denoted by the vectors
\[{\bf A}_{\rm M}^{(n)}=\left(
\begin{array}{c} A_{\rm M}^{(n)}(1,-1) \\ A_{\rm M}^{(n)}(1,0) \\
  \vdots \\ A_{\rm M}^{(n)}(l,m) \\ A_{\rm M}^{(n)}(l,m+1) \\ \vdots \\
  A_{\rm M}^{(n)}(l+1,m) \\ \vdots \\ A_{\rm M}^{(n)}(l_{\rm
    max},l_{\rm max}) \end{array}\right)\;,\quad
{\bf a}^{(n)}=\left(
\begin{array}{c} a^{(n)}_{1,-1} \\ a^{(n)}_{1,0} \\
  \vdots \\ a^{(n)}_{l,m} \\ a^{(n)}_{l,m+1} \\ \vdots \\
  a^{(n)}_{l+1,m} \\ \vdots \\ a^{(n)}_{l_{\rm
    max},l_{\rm max}} \end{array}\right)\;,\quad\]
where the coefficients in the vectors ${\bf A}^{(n)}_{\rm E}$, ${\bf
  B}^{(n)}_{\rm M}$ and ${\bf B}^{(n)}_{\rm E}$ are ordered in analogy
to ${\bf A}^{(n)}_{\rm M}$. The matrix elements can be determined by
using the expression for the
curl terms in Eq.~\ref{EReihen} with $r$ depending on $\theta$ and $\varphi$
\begin{eqnarray}
  \label{NablaxfX}
\vec{\nabla}\times f_l(kr){\bf
  X}_{l,m}&=&\frac{1}{r}\frac{\partial}{\partial
  r}\left[rf_l(kr)\right]{\bf e}_{\rm r}\times{\bf X}_{l,m}+{\bf
  e}_{\rm r}\frac{\rm
  i}{r}\left[\sqrt{l(l+1)}f_l(kr)Y_{l,m}\right.\nonumber\\
& &-\left.\frac{1}{\sqrt{l(l+1)}}\frac{\partial f_l(kr)}{\partial
  r}\left(\frac{\partial r}{\partial\theta}\frac{\partial
    Y_{l,m}}{\partial\theta}+\frac{1}{\sin^2\theta}\frac{\partial
    r}{\partial\varphi}\frac{\partial
    Y_{l,m}}{\partial\varphi}\right)\right]\;.
\end{eqnarray}
Then the matrixelements get the form
\begin{eqnarray}
  \label{Mr1}
  {\sf M}^1_{\rm r}\left(l,m;l',m'\right)&=&\int{\rm d}\Omega\,
  h_l^{(1)}(kr)\cdot\frac{\rm
    i}{\sqrt{l(l+1)}}\left[-{\rm n}_\theta\frac{\partial
      Y_{l,m}}{\partial\theta}-{\rm
      n}_\varphi\frac{1}{\sin\theta}\frac{\partial
      Y_{l,m}}{\partial\varphi}\right]\cdot Y^*_{l',m'}\;,\\
  \label{Mr2}
  {\sf M}_r^2\left(l,m;l',m'\right)&=&\int{\rm
    d}\Omega\,\frac{1}{r}\frac{\partial}{\partial
    r}\left[r\,h_l^{(1)}(kr)\right]\frac{\rm
    i}{\sqrt{l(l+1)}}\left[{\rm n}_\theta\frac{1}{\sin\theta}\frac{\partial
      Y_{l,m}}{\partial\varphi}-{\rm n}_\varphi\frac{\partial
      Y_{l,m}}{\partial\theta}\right]\cdot Y^*_{l',m'}\;,\\
  \label{Mt1}
  {\sf M}_\theta^1\left(l,m;l',m'\right)&=&\int{\rm
    d}\Omega\,h_l^{(1)}(kr){\rm n}_{\rm r}\frac{\rm
    i}{\sqrt{l(l+1)}}\frac{\partial Y_{l,m}}{\partial\theta}\cdot
  Y_{l',m'}^*\;,\\
  {\sf M}^2_\theta\left(l,m;l',m'\right)&=&\int{\rm d}\Omega\frac{\rm
      i}{r}\left[{\rm
      n}_\varphi\left(\sqrt{l(l+1)}h_l^{(1)}(kr)Y_{l,m}-\frac{\partial
      h^{(1)}_l(kr)}{\partial r}\left(\frac{\partial
      r}{\partial\theta}\frac{\partial
      Y_{l,m}}{\partial\theta}+\frac{1}{\sin^2\theta}\frac{\partial
      r}{\partial\varphi}\frac{\partial
      Y_{l,m}}{\partial\varphi}\right)\right)\right.\nonumber\\
& &\qquad\qquad-\left.{\rm n}_{\rm
  r}\frac{1}{\sqrt{l(l+1)}}\frac{\partial}{\partial
  r}\left[r\,h_l^{(1)}(kr)\right]\frac{1}{\sin\theta}\frac{\partial
  Y_{l,m}}{\partial\theta}\right]\cdot Y^*_{l',m'}\;,\\
  \label{Mt2}
  \label{Mp1}
  {\sf M}^1_\varphi\left(l,m;l',m'\right)&=&\int{\rm
    d}\Omega\,h_l^{(1)}(kr){\rm
    n}_r\frac{i}{\sqrt{l(l+1)}}\frac{1}{\sin\theta}\frac{\partial
    Y_{l,m}}{\partial\varphi}\cdot Y^*_{l',m'}\;,\\
  {\sf M}^2_\varphi\left(l,m;l',m'\right)&=&\int{\rm d}\Omega\frac{\rm
    i}{r}\left[{\rm n}_{\rm
      r}\frac{1}{\sqrt{l(l+1)}}\frac{\partial}{\partial
      r}\left[rh_l^{(1)}(kr)\right]\frac{\partial
      Y_{l,m}}{\partial\theta}-{\rm
      n}_\varphi\left(\sqrt{l(l+1)}\,h_l^{(1)}(kr)Y_{l,m}\right.\right.
\nonumber\\
& &\qquad\qquad-\left.\left.\frac{1}{\sqrt{l(l+1)}}\frac{\partial
  h_l^{(1)}(kr)}{\partial r}\left(\frac{\partial
    r}{\partial\theta}\frac{\partial
    Y_{l,m}}{\partial\theta}+\frac{1}{\sin^2\theta}\frac{\partial
    r}{\partial\varphi}\frac{\partial
    Y_{l,m}}{\partial\varphi}\right)\right)\right]\cdot Y^*_{l',m'}\;,\\
  \label{Mp2}
  \label{S1}
  {\sf S}^1\left(l,m;l',m'\right)&=&\int{\rm d}\Omega\frac{\rm
    i}{\sqrt{l(l+1)}}\left[{\rm
      n}_\theta\frac{1}{\sin\theta}\frac{\partial
      Y_{l,m}}{\partial\varphi}-{\rm n}_\varphi\frac{\partial
      Y_{l,m}}{\partial\theta}\right]\cdot Y_{l',m'}^*\;,\\
  \label{S2}
  {\sf S}^2\left(l,m;l',m'\right)&=&\int{\rm d}\Omega\frac{\rm
    i}{r}\left[{\rm n}_{\rm
      r}\left(\sqrt{l(l+1)}h_l^{(1)}(kr)Y_{l,m}-\frac{\partial
      h_l^{(1)}(kr)}{\partial
      r}\frac{1}{\sqrt{l(l+1)}}\left(\frac{\partial
      r}{\partial\theta}\frac{\partial
      Y_{l,m}}{\partial\theta}\right.\right.\right.\nonumber\\
& &\qquad\qquad+\left.\left.\left.\frac{1}{\sin^2\theta}\frac{\partial
        r}{\partial\varphi}\frac{\partial
        Y_{l,m}}{\partial\varphi}\right)\right)+\frac{\partial}{\partial
    r}\left[rh_l^{(1)}(kr)\right]\frac{1}{\sqrt{l(l+1)}}\left({\rm
    n}_\theta\frac{\partial Y_{l,m}}{\partial\theta}\right.\right.\nonumber\\
& &\qquad\qquad+\left.\left.\frac{{\rm
        n}_\varphi}{\sin\theta}\frac{\partial
      Y_{l,m}}{\partial\varphi}\right)\right]\cdot Y_{l',m'}^*\;.
\end{eqnarray}
Replacing the Hankel functions $h_l^{(1)}(kr)$ by the Bessel functions
$j_l(k_1r)$ the matrix elements for the matrices with the suffix 3
and 4 can be derived (see Eq.~\ref{EReihen}). Using as many $(l',m')$
-combinations as
$(l,m)$-combinations all matrices will be quadratic. Thus
Eq.~\ref{MatrixGl} becomes a matrix equation off the form
\begin{equation}
  \label{compact}
  \left(\begin{array}{cccc}{\sf S}^1 & {\sf S}^2 & {\sf S}^3 & {\sf
        S}^4 \\ {\sf M}_{\rm r}^1 & {\sf M}_{\rm r}^2 & {\sf M}_{\rm
        r}^3 & {\sf M}_{\rm r}^4 \\ {\sf M}_\theta^1 & {\sf
        M}_\theta^2 & {\sf M}_\theta^3 & {\sf M}_\theta^4 \\ {\sf
        M}_\varphi^1 & {\sf M}_\varphi^2 & {\sf M}_\varphi^3 & {\sf
        M}_\varphi^4 \end{array}\right)
  \left(\begin{array}{c} {\bf A}_{\rm M}^{(n)} \\ {\bf A}_{\rm E}^{(n)} \\ {\bf
    B}_{\rm M}^{(n)} \\ {\bf B}_{\rm E}^{(n)} \end{array}\right) =
  \left(\begin{array}{c} 2\pi{\bf a}^{(n)} \\ 0 \\ 0 \\ 0 \end{array}\right)\;.
\end{equation}
which can be solved by standard numerical techniques. Moreover it is
only necessary to calculate the
coefficients $A_{\rm E}^{(n)}$ and $A^{(n)}_{\rm M}$ since they fully
determine the $n$-th harmonic radiated field. This completes the
solution of the nonlinear Mie scattering problem for arbitrary cluster
shapes and arbitrary complex index of refraction.

Note that the matrix equations and the many equations presented in
this chapter are necessary to describe the interesting physics. They
are a consequence of the nonsphericity of the
particles implying boundary conditions that vary on the surface of the
particle. This leads to the coupling of different
electromagnetic modes. This is in contrast to the radiation by a
spherical cluster. This
feature is of general
validity independent of the approximations made for the sources. One
should study experimentally this conclusion of the theory. In
the case of spherical particles all equations will collapse to those
of the spherical case.
\section{discussion}
First we discuss the special case of a spherical cluster where no
magnetic multipoles are
radiated and thus the coefficients $A^{(n)}_{\rm M}$ and $B^{(n)}_{\rm M}$ are
equal to zero. Furthermore the orthogonality of the spherical
harmonics
\begin{equation}
\int
Y_{l,m}\left(\theta,\phi\right)Y^*_{l',m'}\left(\theta,\phi\right)d\Omega=
\delta_{l,l'}\delta_{m,m'}\;,
\label{YY*}
\end{equation}
can be applied in the integrals of
Eq.~\ref{Mr1}--\ref{S2} since the radius $r$ and the normal vector
{\bf n} are independent of $\theta$ and $\varphi$. Thus the matrices
${\sf M}_{\rm r}^i$ and ${\sf S}^{1,3}$ become equal to zero and the
other matrices become diagonal. The coefficients $A_{\rm
  E}^{(n)}(l,m)$ can then directly obtained from $a_{l,m}^{(n)}$.

Thus deviations from the spherical shape cause that one mode of the
source can excite several modes of the field which is expressed by the
fact that the coefficients of the fields are coupled to the coefficients of the
source by a system of equation. The shape of the particle directly governs
the integrals in Eqs.~\ref{Mr1}--\ref{S2} and their nontrivial
form for nonspherical particles leads to the system of equations in
Eq.~\ref{MatrixGl} which are valid for arbitrary cluster shapes. Also no
restrictions to the size of the complex index of refraction are
necessary in contrast to theories where a value of $|\epsilon|\approx
1$ is needed like the Rayleigh-Gans theory.

Since our theory needs results from theories for the linear scattering it
already includes all their numerical problems. In principle all modes
must be summed in the series
in Eq.~\ref{Eent} to represent nonspherical particles
by spherical coordinates. So one limitation of the theory is the
finite number of $(l,m)$-combinations which can be taken into
account. As a consequence the matrices {\sf M} and
{\sf S} in Eq.~\ref{MatrixGl} will become very large for strong
deviations from the spherical shape. This purely numerical limitation
is natural for particle shapes which cannot be described by coordinates
which decouple the Helmholtz equation.

On the other hand the theory has the advantage of a very compact form
since the source for the nonlinear response is approximated just by
$\sigma^{(n)}=(\sigma^{(1)})^n$ according to \"Ostling et
al.~\cite{Sta} and it
only needs the boundary conditions for the electric fields from
Eqs.~\ref{nD=sigma} and \ref{nxE=0} for the derivation of the
radiated fields from the source.
The theory is especially valid for
particles with sizes in the Mie range since
it completely takes into account the combinations of the multipoles in the
higher harmonic case (see Eq.~\ref{al2}) which are characteristic
for the radiation of a particle in the Mie range. The
approximation $\sigma^{(n)}=(\sigma^{(1)})^n$ for the source term
should be critically accessed if one
is interested in details of the scattering profiles and effects due to
the geometry of the cluster.

One of the main purposes of this paper was to stimulate again work on
Mie scattering from nonspherical particles. Clearly
further work is necessary to demonstrate the validity of the analysis
presented here. Our theory might stimulate the mathematical treatment
of Mie scattering in the
nonlinear case.

\end{document}